\documentclass[10pt,letterpaper]{article}
\usepackage[top=0.85in,left=2.75in,footskip=0.75in,marginparwidth=2in]{geometry}

\usepackage[utf8]{inputenc}

\usepackage{cite}

\usepackage{nameref,hyperref}

\usepackage[right]{lineno}

\usepackage{microtype}
\DisableLigatures[f]{encoding = *, family = * }

\raggedright
\usepackage{changepage}

\usepackage[aboveskip=1pt,labelfont=bf,labelsep=period,singlelinecheck=off]{caption}

\makeatletter
\renewcommand{\@biblabel}[1]{\quad#1.}
\makeatother

\usepackage{lastpage,fancyhdr,graphicx}
\usepackage{epstopdf}
\fancyheadoffset[L]{2.25in}
\fancyfootoffset[L]{2.25in}

\usepackage{color}

\definecolor{Gray}{gray}{.25}

\usepackage{graphicx}

\usepackage{sidecap}

\usepackage{wrapfig}
\usepackage[pscoord]{eso-pic}
\usepackage[fulladjust]{marginnote}
\reversemarginpar

\usepackage{multirow}
\usepackage{afterpage}

\begin{document}
\vspace*{0.35in}

% title goes here:
\begin{flushleft}
{\Large
\textbf\newline{Interpretable deep learning in single-cell omics}
}
\newline
% authors go here:
\\
Manoj M Wagle\textsuperscript{1,2,3,+},
Siqu Long\textsuperscript{1,3,+},
Carissa Chen\textsuperscript{1,3},
Chunlei Liu\textsuperscript{1,3},
Pengyi Yang\textsuperscript{1,2,3,4,*}
\\
\bigskip
\bf{1} Computational Systems Biology Unit, Children's Medical Research Institute, Faculty of Medicine and Health, The University of Sydney, Westmead, NSW 2145, Australia
\\
\bf{2} School of Mathematics and Statistics, Faculty of Science, The University of Sydney, Camperdown, NSW 2006, Australia
\\
\bf{3} Sydney Precision Data Science Centre, The University of Sydney, Camperdown, NSW 2006, Australia
\\
\bf{4} Charles Perkins Centre, The University of Sydney, Camperdown, NSW 2006, Australia
\\
\bf{+} Equal contribution
\\
\bigskip
* Correspondence: pengyi.yang@sydney.edu.au

\end{flushleft}

\section*{Abstract}

Recent developments in single-cell omics technologies have enabled the quantification of molecular profiles in individual cells at an unparalleled resolution. Deep learning, a rapidly evolving sub-field of machine learning, has instilled a significant interest in single-cell omics research due to its remarkable success in analysing heterogeneous high-dimensional single-cell omics data. Nevertheless, the inherent multi-layer nonlinear architecture of deep learning models often makes them `black boxes' as the reasoning behind predictions is often unknown and not transparent to the user. This has stimulated an increasing body of research for addressing the lack of interpretability in deep learning models, especially in single-cell omics data analyses, where the identification and understanding of molecular regulators are crucial for interpreting model predictions and directing downstream experimental validations. In this work, we introduce the basics of single-cell omics technologies and the concept of interpretable deep learning. This is followed by a review of the recent interpretable deep learning models applied to various single-cell omics research. Lastly, we highlight the current limitations and discuss potential future directions. We anticipate this review to bring together the single-cell and machine learning research communities to foster future development and application of interpretable deep learning in single-cell omics research.
% now start line numbers
%\linenumbers

\section*{Introduction}

The advances in high-throughput omics technologies have transformed our ability to probe molecular programs at a large scale, providing insight into the complex mechanisms underlying various biological systems and diseases. Until recently, most early omics technologies have been typically applied to profile a population of cells, known as `bulk' profiling \cite{hasin2017multi}, where the heterogeneity of cells and cell types are masked by the average signal across the cell population \cite{wang2010single}. Recent establishments of technologies such as single-cell RNA-sequencing (scRNA-seq) \cite{tang2009mrna,shalek2013single} and single-cell assay for transposase-accessible chromatin by sequencing (scATAC-seq) \cite{buenrostro2015single} enables the dissection of cellular composition and heterogeneity at the single-cell level based on their gene expression and chromatin accessibility profiles. The latest advancements in single-cell omics technologies towards multimodality have further made it possible to obtain multimodal measurements simultaneously from the same cell in a single experiment \cite{baysoy2023technological}. These new developments in the single-cell omics field hold great promise for unlocking genetic information at an unparalleled resolution for understanding multi-layered molecular networks that underlie a broad range of cellular processes and diseases \cite{badia2023gene,kim2023gene}.\\

Deep learning, a rapidly evolving sub-field of machine learning, has gained considerable attention in the single-cell community for its capability to deal with heterogeneous, sparse, noisy, and high-dimensional single-cell omics data and versatility in handling a wide range of applications \cite{ma2022deep}. For example, deep learning models have been demonstrated to excel in tasks such as dimension reduction, batch effect removal, data imputation, cell type annotation, and inferring cellular trajectories \cite{li2020deep, tian2019clustering, arisdakessian2019deepimpute, yu2023ensemble, li2023sctour}. Nevertheless, deep learning models are well known for their lack of interpretability \cite{von2021transparency}. That is, predictions made by these models are often hard to interpret, especially towards understanding the underlying molecular mechanisms that drive cellular processes and phenotype. To this end, improving model interpretability has attracted increasing attention, in particular, in applications such as identifying molecular regulators and reconstructing molecular networks \cite{huang2023evaluation,chen2023transformer,fortelny2020knowledge,lotfollahi2023biologically}.\\

In this work, we review the basics of single-cell omics technologies and key principles behind interpretable deep learning. We next summarise the latest development of interpretable deep learning models specifically tailored to single-cell omics research, providing a global view of the current applications in the main interpretable deep learning taxonomy. Finally, we discuss the challenges and potential future directions in this burgeoning field. We hope that this review will shed light on the current state of the field and guide researchers toward making deep learning both robust and reliable for single-cell research.

\section*{Fundamentals of single-cell omics and interpretable deep learning}

\subsection*{The advent of single-cell omics technologies}\hfill

\begin{figure}[!th]
    \centering
    \includegraphics[width=0.8\textwidth]{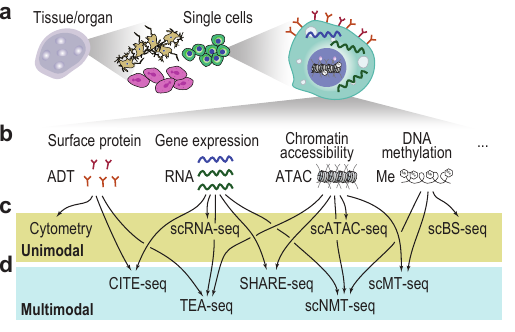}
    \caption{Summary illustration of single-cell omics. (a) A schematic of single cells from different complex tissues/organs. (b) Molecular attributes in a single cell and their corresponding modalities in single-cell omics. (c, d) Example unimodal (c) and multimodal (d) single-cell omics technologies.}\label{fig1}
\end{figure}

The establishment of scRNA-seq technologies that enable transcriptomic profiling at single-cell resolution (Fig. \ref{fig1}a) has revolutionised biomedical research and has since emerged as a powerful tool for dissecting cellular composition \cite{tang2009mrna}. With its potential to reveal variability in cell-to-cell gene expression at an unparalleled accuracy, the use of scRNA-seq has led to ground-breaking discoveries that are unattainable from bulk data, such as cell type annotation for model organisms \cite{iram2018single,the2022tabula} and cell lineage tracing during development and disease progression \cite{kester2018single,wagner2020lineage}. The development of single-cell techniques that profile other modalities (e.g., scATAC-seq (Buenrostro et al., 2015) and single-cell bisulfite sequencing [scBS-seq] of methylomes \cite{smallwood2014single}), and their combination with other single-cell techniques such as cytometry \cite{spitzer2016mass} have led to the generation of additional data modalities and together can provide a more holistic view of the multi-layered molecular programs in single cells (Fig. \ref{fig1}b,c). Nevertheless, these `unimodal' single-cell omics technologies are often independently applied, and each molecular attribute is profiled separately, creating significant difficulties in data modality integration from such `unpaired' data.\\

The recent advance of single-cell omics technologies towards multimodality alleviates the difficulties in integrating data modalities from unpaired data generated by unimodal technologies by measuring multiple data modalities from each single cell \cite{baysoy2023technological}. Thus, the multimodal single-cell omics data generated from such `paired' experiments provide a true single-cell view across multiple molecular attributes. For example, cellular indexing of transcriptomes and epitopes by sequencing (CITE-Seq) \cite{stoeckius2017simultaneous}, simultaneous high-throughput ATAC and RNA expression with sequencing (SHARE-seq) \cite{ma2020chromatin}, and simultaneous single-cell methylome and transcriptome sequencing (scMT-seq) \cite{hu2016simultaneous} each capture a different combination of two modalities in a single cell (Fig. \ref{fig1}d), and techniques such as TEA-seq \cite{swanson2021simultaneous} and single-cell nucleosome, methylation and transcription sequencing (scNMT-seq) \cite{clark2018scnmt} can capture a combination of three modalities in a single cell. A recent review has summarised a comprehensive list of multimodal single-cell omics technologies \cite{vandereyken2023methods}, and the integrative analyses of such data are poised to revolutionise molecular and cellular biology by transforming our understanding of molecular regulators and networks that underlie a broad range of cellular processes and diseases.

\subsection*{Interpretability of deep learning models}\hfill

While the term `model interpretability' still lacks a universal consensus of definition in the machine learning community \cite{lipton2018mythos}, it is broadly considered to be the ability of the model to generate, explain, and present in human understandable terms its decision-making process or insights of data \cite{doshi2017towards,murdoch2019definitions,allen2023interpretable}. Deep learning models, while incredibly successful in their application to various domains, are typically considered as `black boxes' for the lack of interpretability \cite{rudin2019stop} due to their complex multi-layer non-linear architecture, activation functions, and potentially a large number of parameters \cite{zhang2021survey}. Such opacity poses significant challenges in establishing trust and ensuring accurate validation, especially in domains such as molecular biology and biomedicine, where the reasoning behind predictions is crucial for our understanding and subsequent applications \cite{novakovsky2023obtaining}.\\

Adopting popular taxonomies in machine learning \cite{doshi2017towards,murdoch2019definitions,allen2023interpretable}, the interpretability of deep learning models can be largely categorised into (i) {\it intrinsic}, whether prior knowledge or interpretable designs are incorporated into the neural network structures, and {\it post-hoc}, where additional analyses are performed to extract interpretable knowledge from the trained neural networks \cite{murdoch2019definitions}, and (ii) {\it model-specific}, where the interpretability is tailor-made for a specific neural network design, and {\it model-agnostic}, where the techniques could be used across different neural network architectures. For example, post-hoc feature attribution techniques such as Shapley value estimation \cite{NIPS2017_8a20a862} and LIME \cite{ribeiro2016should} are frequently used for identifying important learning features from trained neural networks and are generally considered to be model-agnostic interpretations as they are applicable to various neural network architectures. In contrast, the design of specific model architectures for interpretable learning, such as the use of transformer networks with attention layers, is intrinsic and specific to the model \cite{chefer2021transformer}. In the next section, we will review the current works of interpretable deep learning applications to single-cell omics research in light of these overarching categories.

\section*{Harnessing the power of interpretable deep learning for single-cell omics research} \hfill

\begin{figure}[!th]
    \centering
    \includegraphics[width=0.85\textwidth]{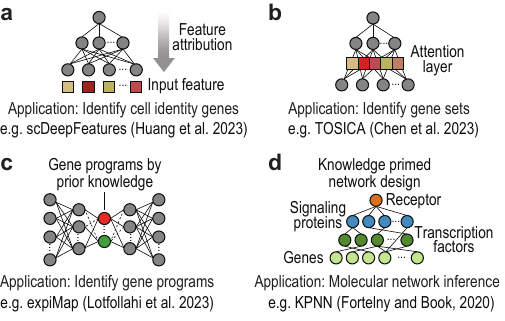}
    \caption{Schematic of example interpretable deep learning models applied to single-cell omics. (a) Using post-hoc feature attribution techniques to identify cell identity genes that distinguish cell types \cite{huang2023evaluation}. Colours denote the estimated importance of input features (b) Designing an intrinsic and model-specific attention layer to detect gene sets for annotating cell types \cite{chen2023transformer}. Colours denote the estimated importance of latent features (c) Incorporating prior knowledge to design embeddings for detecting activated gene programs underlie cell types \cite{lotfollahi2023biologically}. Colours denote different gene programs. (d) Using prior knowledge to design neural network architectures for modelling molecular networks \cite{fortelny2020knowledge}. Colours denote different molecular species.}\label{fig2}
\end{figure}

The application of deep learning models to single-cell omics data analysis has been met with remarkable success owing to their capability to deal with challenging data characteristics (e.g. heterogeneity, sparsity, noise, high-dimensionality) and versatility in handling a wide range of applications in single-cell omics research. Nonetheless, deep learning models often lack interpretability, complicating efforts towards understanding the underlying molecular mechanisms that drive cellular processes and phenotype. In recognition of this limitation, increasing research has been directed to interpretable deep learning in single-cell omics. This section summarises the latest developments in this fast-moving field based on the biological applications of model interpretability, such as identifying cell identity genes and molecular features (Fig. \ref{fig2}a), discovering gene sets (Fig. \ref{fig2}b) or gene programs (iii) (Fig. \ref{fig2}c) that underlie cell types, and inferring molecular networks (Fig. \ref{fig2}d) among other applications.

\subsection*{Identifying cell identity genes and molecular features from unimodal data}\hfill

The identification of genes and other molecular features that underlie cell identity and can discriminate cell types is an essential task in single-cell omics data analysis \cite{yang2021feature}. scDeepFeatures exemplifies the use of several post-hoc approaches for identifying cell identity genes from scRNA-seq data, where a simple multilayer perceptron (MLP)-based neural network is trained to classify cell types, and subsequently various feature attribution techniques (e.g. LIME, feature ablation, occlusion, DeepLIFT) are applied to identify genes that discriminate cell types \cite{huang2023evaluation}. Similarly, Hu et al. proposed a post-hoc and model-agnostic data permutation approach to extract surface proteins from cytometry data using a convolutional neural network \cite{hu2020robust}. Alternatively, scMGCA is a post-hoc and model-specific approach where a graph convolutional autoencoder is first used to learn an embeddings-by-cells matrix from the input data of a normalised count matrix and a cell graph, and a post-hoc embedding analysis procedure is used to identify key cell identity genes that separate cell types and are functionally enriched in the Gene Ontology (GO) analysis \cite{yu2023topological}. scETM serves as a good example for intrinsic and model-specific approaches and uses a variational autoencoder and a linear decoder to factorise the input data into a tri-factor of cells-by-topics, topics-by-embeddings, and embeddings-by-genes matrices. It allows the incorporation of prior pathway information and together enables the identification of interpretable gene markers and cellular signatures when integrating multiple scRNA-seq datasets \cite{yang2022scbert}. scBERT is another intrinsic and model-specific approach that uses a transformer with an attention layer to encode the pre-training data into an embeddings-by-cells matrix and leverage this embedding matrix for subsequent cell-type annotation of future datasets. Cell-type discriminative genes and their long-range dependencies are captured by the attention layer in the model \cite{yang2022scbert}. Finally, siVAE uses cell-wise and gene-wise variational autoencoders to learn interpretable embeddings linked to important genes and co-expression networks \cite{choi2023sivae}.

\subsection*{Detecting genes and molecular features from multimodal data}\hfill

Similar to gene selection from unimodal scRNA-seq data, various methods have also been developed for identifying other molecular features from multimodal single-cell omics data. To this end, several popular approaches make use of autoencoders where multiple data modalities are integrated through embedding learning. For example, Matilda uses a multi-task learning framework of a variational autoencoder and a classification head to classify cell types and a post-hoc feature attribution procedure to identify molecular features from multimodal single-cell omics data that contribute to the classification of each cell type \cite{liu2023multi}. Similarly, UnitedNet uses a dual autoencoder framework for integrating data modalities and subsequently identifying important molecular features using a post-hoc feature attribution technique of Shapley value estimation \cite{tang2023explainable}. totalVI also learns from multiple data modalities using a variational autoencoder and offers a post-hoc archetypal analysis to interpret each latent dimension by relating them to the molecular features in the input data \cite{gayoso2021joint}. Lastly, scMM builds a mixture-of-experts of variational autoencoders for integrating data modalities and then performs a post-hoc step to traverse latent dimensions for identifying molecular features strongly associated with each latent dimension \cite{minoura2021mixture}. These four methods can be generally considered model-agnostic given that the post-hoc approaches they used for identifying molecular features, albeit different, are largely model-independent.

\subsection*{Discovering gene sets and programs that govern cell identity and states}\hfill

Another essential task that extends to gene identification is to discover gene sets or gene programs that jointly govern cell identity, cell states, and cellular processes. scDeepSort represents one of the first methods that use GNNExplainer, a post-hoc and model-agnostic approach, for selecting gene sets that are predictive of cell types from a weighted graph neural network trained on scRNA-seq data \cite{shao2021scdeepsort}. Compared to post-hoc and model-agnostic approaches, methods that rely on intrinsic and model-specific mechanisms have enjoyed more popularity. For example, scCapsNet uses a capsule network model for interpretable learning and captures gene sets that are predictive to cell types from scRNA-seq data \cite{wang2020interpretable}. TOSICA uses a multi-head self-attention network to incorporate prior biological knowledge for identifying gene sets that belong to pathways or regulons for cell type annotation using scRNA-seq data \cite{chen2023transformer}. Alternatively, VEGA utilises prior knowledge of gene modules for designing an interpretable latent space in variational autoencoders and subsequently detecting active modules from models trained on scRNA-seq data \cite{seninge2021vega}. Likewise, ExpiMap incorporates prior knowledge of gene programs into a sparsely connected variational autoencoder during pre-training on large scRNA-seq reference atlases and subsequently identifies gene programs that are associated with cell types and cell states in query datasets \cite{lotfollahi2023biologically}. Finally, pmVAE trains multiple variational autoencoders, each incorporating prior information of a pathway module for detecting biological effects such as cell stimulation from scRNA-seq data \cite{gut2021pmvae}.

\subsection*{Inferring molecular networks from single-cell omics data}\hfill

Building on the concept of gene sets and pathways, the next related task is to infer molecular networks, such as gene regulatory networks (GRNs), or capture regulatory relationships among transcription factors (TFs) and their target genes \cite{badia2023gene,kim2023gene}. One of the first methods towards achieving this aim is the knowledge-primed neural network (KPNN), an intrinsic and model-specific approach that explores the design of a sparsely connected neural network based on prior knowledge of genome-wide regulatory networks and subsequently trains the model using scRNA-seq data to learn regulatory strengths as weights of network edges between TFs and their target genes \cite{fortelny2020knowledge}. Alternatively, scGeneRAI attempts to infer GRNs by predicting the expression of a gene from a set of other genes using scRNA-seq data and layer-wise relevance propagation (LRP), a post-hoc and model-specific feature attribution technique \cite{keyl2023single}. Similar to scGeneRAI, STGRNS aims to reconstruct GRNs by predicting TF expressions using gene sets but relies on an intrinsic and model-specific approach where a transformer network with a multi-head attention layer is trained on scRNA-seq data \cite{xu2023stgrns}. More recently, methods such as DeepMAPS have been developed for intrinsic and model-specific interpretable learning from multimodal single-cell omics data \cite{ma2023single}. In particular, DeepMAPS uses a graph autoencoder for integrating data modalities and then inserts the trained graph autoencoder into a heterogeneous graph transformer with mutual attention layers for inferring cell-type-specific GRNs.

\subsection*{Predicting transcription factor binding sites and sequence motifs}\hfill

In GRNs, TFs regulate their target genes through binding to cis-regulatory elements (CREs) that contain specific sequence motifs. The prediction of TF binding sites and sequence motifs therefore goes further than inferring GRNs solely based on gene expression and deepens our understanding of underlying mechanisms of molecular network regulation. Several studies have demonstrated the utility of convolutional neural networks (CNNs) for addressing this task. In particular, ExplainNN predicts TF binding sites and sequence motifs through learning convolutional filters on DNA sequences. The model predictions are validated using scATAC-seq data and curations in databases such as JASPAR \cite{novakovsky2023explainn}. IscPNAM is another CNN-based method for predicting TF binding sites but integrates DNA sequences with bulk data (e.g. ATAC-seq data) for their prediction and generates interpretations from additional attention modules in the network \cite{gong2023interpretable}. Similar to ExplainNN, IscPNAM also uses scATAC-seq data for its prediction evaluation. Lastly, scover also uses CNN for interpretable learning of sequence motifs using convolutional filters. While scATAC-seq data are used for validation purposes in ExplainNN and IscPNAM, they are used for training CNNs in scover for discovering regulatory motifs that are cell-type-specific and reside in distal CREs and therefore can benefit from the cell-type-specific information captured by single-cell omics data \cite{hepkema2023predicting}. Although IscPNAM uses model intrinsic attention modules for interpretations, all three methods rely on intrinsic and model-specific mechanisms for interpretable learning.

\subsection*{Other applications}\hfill

The above applications of interpretable deep learning mostly centre around some closely related tasks of identifying genes and gene sets and inferring their regulatory relationships from unimodal and multimodal single-cell omics data. Beyond these, a few studies have explored other potential applications. Examples include TAPE and UCDBase for bulk transcriptomic data deconvolution using scRNA-seq data. Specifically, TAPE implements a training stage for an autoencoder and an adaptive learning stage to extract interpretable information from the trained model, including cell-type-specific signature matrix and gene expression profiles, and predicted cell-type composition in the bulk data \cite{chen2022deep}. On the other hand, UCDBase first pre-trains a densely-connected neural network model using large-scale scRNA-seq atlases and transfers the model for bulk data deconvolution. The model interpretations are generated using a post-hoc model-specific feature attribution technique of integrated gradients \cite{charytonowicz2023interpretable}. Another example is PAUSE which models transcriptomic variation by using a biologically constrained autoencoder to attribute variations in scRNA-seq data to pathway modules \cite{janizek2023pause}.

\section*{Current challenges and future opportunities} \hfill

Interpretable deep learning has made a significant impact on the single-cell omics research field. However, the current application of interpretable deep learning techniques to single-cell omics research is still limited to a few related tasks of identifying genes and programs and inferring molecular networks they form. Other common tasks, such as trajectory inferences and cell-cell interactions, are crucial in single-cell omics data analysis \cite{heumos2023best} but remain less explored in the context of interpretable deep learning. We anticipate that future method development and application will investigate the potential utility of interpretable deep learning in addressing these tasks.\\

Besides multimodality, the recent development of single-cell omics research is increasingly towards multi-sample and multi-condition. While methods such as ExpiMap have been designed for data generated from different samples and with various perturbations and conditions in mind \cite{lotfollahi2023biologically}, there is still a lack of application of interpretable deep learning models to specifically address these emerging data structures that go beyond typical tasks of cell type annotation, cell identity gene selection, and GRN inference from a normal sample without experimental perturbations. Given the increasing application of single-cell omics techniques to studying human diseases and drug perturbations, we expect to see new methods developed to extract interpretable information from these more complex data structures.\\

Another aspect of single-cell omics technological advancements is in the field of spatial transcriptomics \cite{rao2021exploring}. Few interpretable deep learning methods have been specifically designed to take advantage of the extra spatial information besides the gene expression profile of cells. Yet, such information can be valuable for studying cell-cell communications and cellular microenvironments that drive normal development and diseases such as cancer. Therefore, developing deep learning models to integrate spatial information and other omic data modalities in single cells for interpretable learning may lead to a better understanding of developmental processes and improved treatments of cancer.\\

It is important to note that annotating interpretable deep learning models reviewed in this work into the four overarching categories is useful for their summarisation. However, this is only intended to serve as a conceptual framework for ease of understanding the main strategies used in each method. Some methods use multiple interpretable learning techniques that can fit into more than one category, while others develop new strategies that may not precisely fit into any of these categories. Furthermore, there are additional taxonomy strategies, such as classifying model interpretability to be {\it global} and {\it local} \cite{allen2023interpretable}, but are excluded due to their less utility in summarising methods reviewed in this work.\\

Related to the above, the concept and definition of interpretability are fast-evolving. For example, methods that use simpler neural network architectures and perform linear transformation can be viewed as more interpretable. Besides improving model interpretability, methods that generate biologically meaningful results can also be viewed as more interpretable. For instance, in scvis, increased interpretability is defined as generating embeddings that better preserve the local and global neighbour structures in the original high-dimensional scRNA-seq data \cite{ding2018interpretable}. While these definitions of interpretability are beyond the scope of this review, they are nonetheless important aspects that will contribute to the future development of interpretable deep learning in the single-cell omics research field.

\section*{Conclusions}

Deep learning models, previously viewed as `black boxes' for their lack of interpretability, have become increasingly interpretable due to the recent progress made in interpretable deep learning. This has stimulated a growing interest in using interpretable deep learning techniques for single-cell omics research, as the ability to identify and understand molecular regulators and networks is critical for guiding downstream experimental validations. Here, we briefly introduce the key concepts in single-cell omics technologies and interpretable deep learning techniques and then review the recent advancement in the development and application of interpretable deep learning models to various single-cell omics data analysis tasks. We discuss current challenges and opportunities and hope this review will catalyse this multidisciplinary research field for future development and application of interpretable deep learning to accelerate single-cell omics research.

\subsection*{Acknowledgements} \hfill

P.Y. is supported by a National Health and Medical Research Council Investigator Grant (1173469). The authors thank the useful feedback from the members of the Computational Systems Biology Unit.

\subsection*{Author contributions} \hfill

P.Y. conceptualised this work and supervised all authors to review the literature, and write and edit the manuscript.

\subsection*{Competing interests} \hfill

The authors declare no competing interests.

\nolinenumbers
%\section*{Bibliography}
%This is where your bibliography is generated. Make sure that your .bib file is actually called library.bib
\bibliography{main}

%This defines the bibliographies style. Search online for a list of available styles.
\bibliographystyle{abbrv}

\end{document}